\documentclass[prb,twocolumn,showpacs,showkeys,amsmath]{revtex4}
\usepackage{graphics}
\usepackage{graphicx}
\newcommand{\mb}[1]{\boldsymbol{#1}}
\begin{document}

\title{Magnetotransport in Two-Dimensional Electron Systems  with
Spin-Orbit Interaction}

\author{M.~Langenbuch}
\email{michael.langenbuch@physik.uni-regensburg.de}
\author{M.~Suhrke} 
\altaffiliation{Fraunhofer Institut Naturwissenschaftlich-Technische Trendanalysen, Appelsgarten 2, 53879~Euskirchen, Germany}
\author{U.~R\"{o}ssler}

\affiliation{Institut f\"{u}r Theoretische Physik - Universit\"{a}t
Regensburg, 93040~Regensburg, Germany}

\date{\today}

\begin{abstract}
We present magnetotransport calculations for homogeneous 
two-dimensional electron systems including the Rashba spin-orbit
interaction, which  
mixes the spin-eigenstates and leads to a modified fan-chart with
crossing Landau levels. The quantum mechanical Kubo
formula is evaluated by taking into account spin-conserving scatterers 
in an extension of the self-consistent Born approximation that
considers the spin degree of
freedom. The calculated conductivity exhibits besides the well-known
beating in the Shubnikov-de Haas (SdH) oscillations a modulation which 
is due to a suppression of scattering away from the crossing points of 
Landau levels and does not show up in the density of states. This
modulation, surviving even at elevated temperatures when the SdH
oscillations are damped out, could serve to identify spin-orbit
coupling in magnetotransport experiments. Our magnetotransport
calculations are extended also to lateral superlattices and
predictions are made with respect to $1/B$ periodic oscillations in
dependence on carrier density and strength of the spin-orbit coupling.
\end{abstract}



\maketitle

\section{Introduction}
The Rashba spin-orbit coupling \cite{rash60,bych84} 
that exists in systems with axial symmetry, plays a key role in
spintronics  \cite{wolf01} realized with two-dimensional (2D) carriers
in semiconductor heterostructures as it allows to manipulate the spin by
a gate-controlled confinement potential. Spin-orbit (SO)
coupling mixes the spin states and removes the spin degeneracy for
states with finite momentum. Besides the Rashba term caused by the
asymmetry of the confinement, there exists also a SO coupling due to
the inversion asymmetry of the crystalline structure of the
semiconductor bulk material (Dresselhaus term \cite{dres55}). Both
types of SO coupling combine to an anisotropic spin-splitting of 2D
electrons, which has been analyzed by inelastic light scattering
\cite{juss} and play a role also in weak
localization studies  \cite{pikus_pikus,schie02}. The intimate relation
between spin splitting and spin relaxation, well-known for bulk
material \cite{titkov,dp} has found renewed interest for 2D electrons
\cite{golub,jk}, furthered by the possibility to
measure spin-relaxation times with monopolar optical orientation
\cite{ganichev}.  The zero-field spin splitting \cite{lomm88} has to
compete with the Zeeman spin-splitting if a magnetic field is applied
perpendicular to the plane of the 2D electron system. This results in 
a fan chart showing characteristic crossings of Landau levels, which in
magnetotransport data are detected as beating of the Shubnikov de-Haas 
(SdH) oscillations \cite{nitt97,heid98,schae98,hu99}. For hole sytems, 
the two SO coupling mechanisms have been found to be responsible for
anomalous SdH oscillations \cite{wink00,kepp02}. The structural
asymmetry of the confinement can be tuned into a regime where the
Rashba SO coupling dominates over the  bulk inversion asymmetry
\cite{lomm88}. In spite 
of this current 
interest in the Rashba SO coupling and its relevance for spin-related
transport in  two-dimensional electron systems (2DES) it is surprising
that there is so far no rigorous magnetotransport calculation which
takes this coupling  into account.

Here, we  present fully quantum-mechanical
calculations of the magnetoconductivity including the Rashba SO
interaction.  The  calculations are based on the evaluation of the
Kubo formula with an 
extension of the self-consistent Born 
approximation (SCBA) by taking into account the electron spin degree
of freedom. Our results show besides the known beats in the SdH
oscillations an additional modulation connected with the crossing of
Landau levels. This modulation, which is not seen in the density of
states, arises as we take into account spin-conserving 
impurity scattering which is suppressed when the SO coupled states
are not degenerate. It survives even at higher temperatures,
when SdH oscillations have died out, and could serve, if experimentally 
detected, as another fingerprint of SO interaction. In
lateral superlattices \cite{weis91}, where a 2DES is subjected to a
periodic potential, there exist  besides the SdH
oscillations other  $1/B$ periodic magnetotransport
oscillations due to commensurability between cyclotron radius and
lattice constant \cite{weis89_wink89} and due to the formation of a
miniband structure \cite{albr99,deut01}. From our magnetotransport
calculations for lateral superlattices with weak uni-directional (1D)
modulation we  
predict a splitting of these periods due to SO coupling and calculate
their dependence on its strength and carrier density.

The paper is  organized in the following way. After this introduction
we describe in the section \ref{second_section} the energy spectrum and the eigenstates of a 2DES with
Rashba SO 
interaction. For the calculation of the conductivity we simplify the
energy spectum and introduce a constant spin-splitting model. In the
section \ref{third_section} the SCBA is extended by the spin degree of
freedom to describe the scattering of spin-conserving impurities in a
2DES  
with SO interaction. The effects of this extension
are illustrated  for a two level system, where it is shown that for
non-degenerate SO coupled  states 
the scattering efficiency is suppressed. In section
\ref{sec_conductivity} we present
the conductivity and compare the cases with and without SO
coupling. Finally in the section  \ref{fifth_section} we show results
for a
system with SO interaction and a 1D periodic modulation and study in the power spectrum of the magnetoconductivity the evolution of the characteristic periods with increasing SO coupling.

\section{Energy spectrum and the constant spin-orbit coupling model}
\label{second_section}

The Hamiltonian of a 2DES (in the $xy$-plane) realized in 
the lowest subband of a semiconductor heterostructure with
effective mass $m^{*}$, Rashba SO interaction due to the
$z$-confinement with coupling constant 
$\alpha_{z}$, and Zeeman term with effective $g$-factor $g^{*}$, in a
magnetic field $\mathbf{B} = B \hat{\mathbf{e}}_{z}$, is  given by
\begin{equation}
\label{eqn_hamiltonian}
H = \frac{1}{2 m^{*}} \left( {\pi}_{x}^2 + {\pi}_{y}^2 \right)
- \frac{\alpha_{z}}{\hbar} \, \left( \sigma_{x} \pi_{y} - \sigma_{y}
\pi_{x} \right) + \frac{1}{2} g^{*} \mu_{\mathrm{B}} \sigma_{z} B, 
\end{equation}
where ${\pi_{\mu}}$ denotes the kinetic momentum and $\sigma_{\mu}$
the Pauli spin matrices, $\mu \in \{ x,y,z \}$. The spin ({\it
up}/{\it down}) 
 is quantized in $z$-direction. The energy spectrum is isotropic and
without the magnetic field, $B=0$, it depends on the wavevector
$\mb{k}$ and is given by \cite{rash60}
\begin{equation}
E^{\pm}_{\mb{k}} = \frac{\hbar^2 \mb{k}^2}{2 m^{*}} \pm \alpha_{z}
|\mb{k}|. 
\end{equation}
The SO coupling lifts the spin degeneracy even without external
magnetic field  and the energy branches are split by
\begin{equation}
\label{eqn_deltasok}
\Delta_{\mathrm{SO}} = 2 \alpha_{z} |\mb{k}|.
\end{equation}
Including the external magnetic field, the Hamiltonian can be formulated with ladder
operators $a^{\dagger} 
|n\rangle = \sqrt{n+1}\, 
|n+1\rangle$, $a|n\rangle = \sqrt{n} \, |n-1 \rangle $.  In the Pauli
representation the Hamiltonian can be written as
\begin{equation}
\label{eqn_hamiltonian_ladder}
H = \hbar \omega_{c}
\begin{pmatrix}
a^{\dagger} a + \frac{1}{2} + \beta & \alpha a \\
\alpha a^{\dagger} & a^{\dagger} a + \frac{1}{2} - \beta 
\end{pmatrix}.
\end{equation}
with the parameters $\beta = \frac{g^{*} \mu_{B} B}{2 \hbar
\omega_{c}}$  and
$ \alpha = - \frac{ \alpha_{z} \sqrt{2}}{\lambda_{c} \hbar
\omega_{c}}$, the cyclotron frequency $\omega_{c} = \tfrac{e
B}{m^{*}}$ and the magnetic length $\lambda_{c} = \left( \tfrac{\hbar}
{e B} \right)^{1/2}$.

\begin{figure}[h]
\includegraphics[width= \columnwidth]{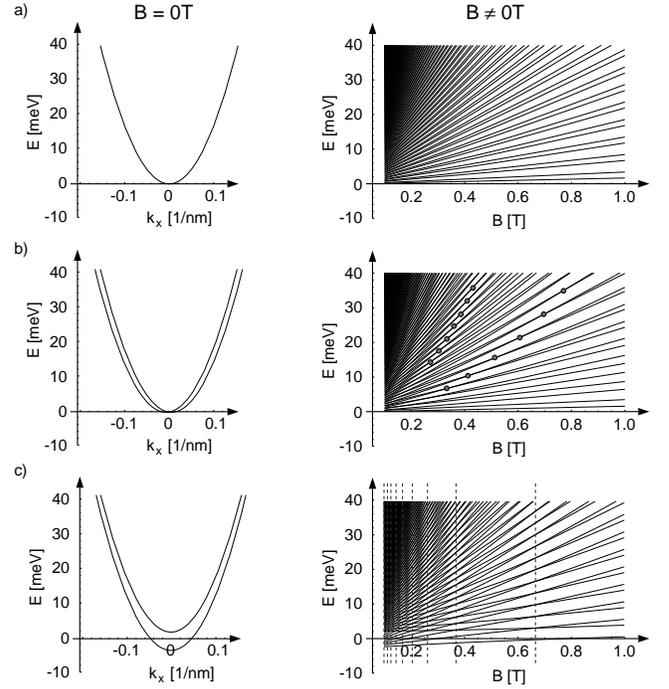}
\caption{\label{fig_one}  Subband Energy dispersion (left) and Landau levels 
(right) of a 2DES (parameters $m^{*} = 0.023 \,m_{0}$, $g^{*} = -14.9$ 
correspond to InAs) a) without SO coupling $\alpha_{z} =
0\,$eVm,  
b) with SO coupling $\alpha_{z}=2.0 \times 10^{-11}\,$eVm, and
c) with constant spin-orbit splitting $\bar{\alpha}_{z} = 2.5 \,$meV.}
\end{figure}
In Fig.~\ref{fig_one} we show the spectrum of Hamiltonian
(\ref{eqn_hamiltonian}) without magnetic field (left) and with magnetic
field (right) for parameter values corresponding to InAs ($m^{*} =
0.023 \,m_{0}$, $g^{*} = -14.9$). In Fig.~\ref{fig_one}a the
situation without SO coupling with spin-degenerate parabolic
dispersion (left) and regular fan-chart of Landau levels (right) is
depicted. Including spin-orbit coupling the picture of
Fig.~\ref{fig_one}b is obtained with the k-dependent splitting of
the  subband dispersion (left) and the characteristic crossing pattern
of 
Landau levels (right). For this calculation the SO coupling parameter
$\alpha_{z}$ was choosen to be $\alpha_{z}=2.0 \times 10^{-11}\,$eVm
close to the experimental values reported for InAs samples
\cite{nitt97}.  
The Hamiltonian (\ref{eqn_hamiltonian_ladder}) indicates that in the
presence of SO coupling the spin states, quantized in $z$-direction and 
used in the Pauli representation, are no longer eigenstates. Instead
Hamiltonian (\ref{eqn_hamiltonian_ladder}) is diagonal for the states 
\cite{rash60}
\begin{equation} 
\label{eqn_soeigenzustaende}
\left| n, r \right\rangle   = 
\begin{pmatrix}
c_{r n}^{u}   \Phi_{n}  \\ c_{r n}^{d}  \Phi_{n+1} 
\end{pmatrix}  
\text{ and }
\left| n, l \right\rangle   = \begin{pmatrix}
c_{l n}^{u}  \Phi_{n-1}  \\ c_{l n}^{d}  \Phi_{n} 
\end{pmatrix}
\end{equation} 
which can be classified by the helicity $\kappa \in \{\mbox{\it
right},\mbox{\it left} \}$ and the Landau index $n = 0, 1, 2,
...\,$. $\Phi_{n}$ are the eigenfunctions of the Landau
oscillator. These {\it right}/{\it left} states evolve with increasing 
SO coupling from the spin {\it up}/{\it down} states, respectively, by
the following choice of the coefficients:
\begin{align}
\label{eqn_coeff_right}
c_{n r}^{u} = \frac{\alpha \sqrt{n+1}}{\sqrt{(n+1) \alpha^2 +
\left( (\tfrac{1}{2} - \beta) - \sqrt{d_{r}} \right)^2}}, \nonumber \\
c_{n r}^{d} = \frac{\left( \tfrac{1}{2} -\beta -
\sqrt{d_{r}}\right)}{\sqrt{(n+1) \alpha^2 +
\left( (\tfrac{1}{2} - \beta) - \sqrt{d_{r}} \right)^2}}, 
\end{align}
for the {\it right} states and 
\begin{align}
\label{eqn_coeff_left}
c_{n l}^{u} = \frac{\alpha \sqrt{n}}{\sqrt{n \alpha^2 +
\left( (\tfrac{1}{2} - \beta) + \sqrt{d_{l}} \right)^2}}, \nonumber \\
c_{n l}^{d} = \frac{\left( \tfrac{1}{2} -\beta +
\sqrt{d_{l}}\right)}{\sqrt{ n \alpha^2 +
\left( (\tfrac{1}{2} - \beta) + \sqrt{d_{l}} \right)^2}}, 
\end{align}
for {\it left} states with $d_{r} = \sqrt{ (n+1) \alpha^2 +
(\frac{1}{2} - \beta)^2}$ and 
$d_{l} = \sqrt{ n \alpha^2 + (\frac{1}{2} - \beta)^2}$. The energy
eigenvalues of these states are
\begin{align}
\label{eqn_eigenvaluesrl}
E_{n r} & = \hbar \omega_{c} \left( 1 + n - \sqrt{ (n+1)   \alpha^2 +
\left(
\tfrac{1}{2} -\beta \right)^2}  \right), \\
E_{n l} & = \;\; \hbar \omega_{c} \;\;\left( n \; + \; \sqrt{ n \alpha^2
+ \left(
\tfrac{1}{2} -\beta
\right)^2} \right).
\end{align}
In the following we use the notation
\begin{equation}
\label{eqn_rechtslinks_basis}
\left| n,\kappa \right\rangle = \sum_{\sigma} c_{n \kappa}^\sigma \left|
n-\tfrac{\sigma-\kappa}{2}, \sigma \right\rangle.
\end{equation}
where $\sigma$ denotes the spin quantized in $z$-direction
(1: {\it up}, -1: {\it down}) and $\kappa$ the helicity (1: {\it
right}, -1: {\it left}).

The low temperature magnetoconductivity is determined by the electron
states close to the Fermi energy. We will be interested mainly in
the low-magnetic field regime where the conductivity is dominated by
contributions   
from Landau levels with $n \gg 1$ and we may simplify the
energy spectrum by replacing 
$\alpha \to      
\bar{\alpha} / \sqrt{a^{\dagger} a + \tfrac{1}{2}}$ with $\bar{\alpha}
= \tfrac{\bar{\alpha}_{z}}{2 \hbar \omega_{c}}$.  With the approximation
$\tfrac{n+1}{n} \to 1$ we arrive at a model with constant
spin-splitting $\Delta_{SO}=2 \bar{\alpha}_{z}$ and the coefficients
of (\ref{eqn_coeff_right}) and (\ref{eqn_coeff_left}) take the forms 
\begin{align}
c_{r}^{u} = \frac{\bar{\alpha}}{\sqrt{ \bar{\alpha}^2 +
\left( (\tfrac{1}{2} - \beta) - \sqrt{\bar{d}} \right)^2}}, \nonumber \\
c_{r}^{d} = \frac{\left( \tfrac{1}{2} -\beta -
\sqrt{\bar{d}}\right)}{\sqrt{ \bar{\alpha}^2 +
\left( (\tfrac{1}{2} - \beta) - \sqrt{\bar{d}} \right)^2}} 
\end{align}
and
\begin{align}
c_{l}^{u} = \frac{\bar{\alpha}}{\sqrt{ \bar{\alpha}^2 +
\left( (\tfrac{1}{2} - \beta) + \sqrt{\bar{d}} \right)^2}},  \nonumber \\
c_{l}^{d} = \frac{\left( \tfrac{1}{2} -\beta +
\sqrt{\bar{d}}\right)}{\sqrt{\bar{\alpha}^2 +
\left( (\tfrac{1}{2} - \beta) + \sqrt{\bar{d}} \right)^2}} 
\end{align}
respectively, where $\bar{d} = \sqrt{\bar{\alpha}^2 + (\frac{1}{2} - \beta)^2}$. For 
this constant SO coupling model the energy eigenvalues are  
\begin{align}
E_{n r} & = \hbar \omega_{c} \left( 1 + n - \sqrt{ \bar{\alpha}^2 +
\left(
\tfrac{1}{2} -\beta \right)^2}  \right), \\
E_{n l} & = \;\;\hbar \omega_{c} \;\;\left( n \; + \; \sqrt{
\bar{\alpha}^2 + \left(
\tfrac{1}{2} -\beta
\right)^2  } \right).
\end{align}
The energy spectrum  of this model (Fig.~\ref{fig_one}c) consists of
the two branches of 
{\it right} and 
{\it left} states  which by our choice of $\bar{\alpha}_{z}$ are
shifted by 
$\Delta_{\mathrm{SO}} = 5\,$meV. The 
crossing of Landau levels takes place  for all levels at
the same  magnetic field. Thus our model preserves  the two main
effects of SO coupling, the crossing of Landau  
levels and the mixing of spin components {\it up}/{\it down}.

\section{SCBA with spin-degree of freedom}
 \label{third_section}

For the calculation of the conductivity scattering has to be taken 
into account. The impurity-averaged Green function of the system  $G$ 
is given by the Dyson equation
\begin{equation}
 G = G_{0} + G_{0} \Sigma G.
\end{equation}
The scattering is included by the selfenergy $\Sigma$ in
self-consistent Born approximation (SCBA)
\cite{ando82}.  The matrixelements of the selfenergy $\langle \alpha |
\Sigma | \alpha' \rangle = \Sigma_{\alpha \alpha'}$, $\alpha= (n
\kappa)$ read in the basis of eigenstates (\ref{eqn_rechtslinks_basis})
\begin{equation}
\Sigma_{\alpha \alpha'} = \sum_{\beta \beta'} \Gamma_{\alpha \beta
\beta' \alpha'} G_{\beta \beta'}.
\end{equation}
The kernel $\Gamma_{\alpha \beta \beta' \alpha'}$ is given by the expression
\begin{equation}
\label{eqn_gamma}
\Gamma_{\alpha \beta \beta' \alpha'} = \int \tfrac{dq^{2}}{(2
  \pi)^{2}} \left| \tilde{v}_{\mathrm I}(\mb{q}) \right|^2 
\left\langle \alpha \right|  e^{i \mb{q} \mb{r}} 
\left|       \beta \right\rangle
\left\langle \beta' \right|  e^{-i \mb{q} \mb{r}} 
\left|       \alpha' \right\rangle,
\end{equation}
where $\tilde{v}_{\mathrm I}(\mb{q})$ denotes the Fourier
transform of the impurity potential \cite{gerh75}.

We consider spin-conserving short range impuritiy scattering. By this
choice we neglect magnetic impurities and SO 
coupling with the scattering center, thus without Rashba SO coupling
scattering is possible
only between states with the same spin-quantum number. In the 
new basis (\ref{eqn_rechtslinks_basis}), including the Rashba term,
scattering between states with different
helicity becomes possible. 
For $\delta$-scatterers
Eq.~(\ref{eqn_gamma}) simplifies to  
\begin{align}
\label{eqn_gamma_mitso}
\Gamma_{\alpha \beta \beta' \alpha'} & = \Gamma^2 \sum_{\sigma \sigma'} 
\, \bigl({c}_{n \kappa}^{\sigma}\bigr)^{*} \, {c}_{m
\tilde{\kappa}}^{\sigma} \, \bigl({c}_{m'
\tilde{\kappa}'}^{\sigma'}\bigr)^{*} \,  {c}_{n' \kappa'}^{\sigma'}
\times 
\\ 
& \times \delta_{n-\tfrac{\sigma-\kappa}{2} , n' -\tfrac{\sigma'
-\kappa'}{2}} \;\; \cdot \;\;
\delta_{m-\tfrac{\sigma-\tilde{\kappa}}{2} , m' -\tfrac{\sigma'
-\tilde{\kappa}'}{2}},
\end{align}
where $\Gamma^2 = \frac{1}{2 \pi} \hbar \omega_{c}
\frac{\hbar}{\tau}$ is connected through the relaxation time $\tau$
with the mobility $\mu = e \tau/m^{\ast}$ in the case without magnetic
field. The 
Kronecker symbols have the form  
$\delta_{n',n-\theta}$ and $\delta_{m',m-\mu}$ with $\theta =
\tfrac{1}{2}(\sigma - \sigma' -\kappa + \kappa') = -2,-1,0,1,2$ and 
$\mu = \tfrac{1}{2}(\sigma - \sigma' -\tilde{\kappa} +
\tilde{\kappa}') = -2,-1,0,1,2$. We restrict ourselves to the diagonal 
approximation in the spacial quantum numbers by considering only
$\theta=\mu=0$. In this approximation the selfenergy reads $\Sigma_{n
n'}^{\kappa \kappa'} = \Sigma^{\kappa \kappa'} \delta_{n n'}$ with
$\Sigma^{\kappa \kappa'}$ independent of $n$, and the Green function
$G_{m m'}^{\kappa \kappa'} = G_{m} ^{\kappa \kappa'} \delta_{m m'}$. 
Thus the SCBA  selfenergy becomes
\begin{equation}
\Sigma^{\kappa \kappa'} = \Gamma^{2} \sum_{m} \sum_{\tilde{\kappa}
  \tilde{\kappa}'}  \alpha_{\tilde{\kappa} \tilde{\kappa}'}^{\kappa \kappa'}  G_{m}^{\tilde{\kappa} \tilde{\kappa}'} 
\end{equation}
with
\begin{equation}
\alpha_{\tilde{\kappa} \tilde{\kappa}'}^{\kappa \kappa'} = \sum_{\sigma
\sigma'} \bigl( c_{\kappa}^{\sigma}\bigr)^{*} c_{\tilde{\kappa}}^{\sigma} 
\bigl( c_{\tilde{\kappa}'}^{\sigma'} \bigr)^{*}  c_{\kappa'}^{\sigma'} \;
\delta_{\sigma-\sigma',\kappa-\kappa'} \,
\delta_{\sigma-\sigma',\tilde{\kappa} - \tilde{\kappa}'} 
\end{equation}
and
\begin{equation}
\label{eqn_green_tautau}
\left( G^{\kappa \kappa'}_{m} \right)^{-1} = \left( G_{\!0\,m}^{\;\;\kappa} \right)^{-1}
\delta_{\kappa \kappa'} - \Sigma^{\kappa \kappa'},
\end{equation}
where $ G_{\!0\,m}^{\;\;\kappa}$ is the Green function of the system
without impurities. Using the abbreviations
$\Sigma^{\kappa \kappa} \equiv \Sigma^{\kappa}$, $G^{\kappa \kappa}_{m}
\equiv G^{\kappa}_{m}$ and $\alpha_{\tilde{\kappa} \tilde{\kappa}'}^{\kappa \kappa} \equiv \alpha_{\kappa
\tilde{\kappa}} \delta_{\tilde{\kappa} \tilde{\kappa}'}$ the
selfenergy can now be calculated from 
\begin{equation}
\label{eqn_sigma_tautau}
\Sigma^{\kappa} = \Gamma^{2} \sum_{m \tilde{\kappa}}
\alpha_{\kappa \tilde{\kappa}} G_{m}^{\tilde{\kappa}}
\;\; \;\;\; \text{with} \;\;\;\;\; \alpha_{\kappa \tilde{\kappa}} = \sum_{\sigma}
\left| c^{\sigma}_{\kappa}  \right|^{2} \left| c^{\sigma}_{\tilde{\kappa}}\right|^{2}.
\end{equation}
The off-diagonal elements of the selfenergy will be neglected. 

The density of states for the {\it left} and
{\it right} components are obtained by the trace over the spectral
funktions $ A^{\kappa}_{m} = - \mbox{Im}\left\{ \tfrac{1}{\pi}
G_{m}^{\kappa} \right\}$. 
\begin{align}
\label{eqn_dos_tau}
D^{\kappa}(E) &= \frac{1}{L_{x}L_{y}} \mbox{Tr}_{m} \{ A^{\kappa}_{m}
\} = 
\nonumber \\
 &= - \frac{1}{(\pi \lambda_{\mathrm{c}} \Gamma)^{2}} \, \sum_{\kappa'}
(\alpha^{-1})_{\kappa \kappa'} \,\mbox{Im}\left\{  \Sigma^{\kappa'}(E) \right\}.
\end{align}
Here $(\alpha^{-1})_{\kappa \kappa'}$ is the inverse of the matrix 
formed from the $\alpha_{\kappa \kappa'}$ of
Eq.~(\ref{eqn_sigma_tautau}) for which $\alpha_{\kappa \kappa} + \alpha_{\kappa (-\kappa)} =
1$. For vanishing SO coupling one has $\alpha_{\kappa \tilde{\kappa}}
\to \delta_{\kappa \tilde{\kappa}}$  and the standard SCBA result is  
reproduced. For strong SO coupling if $\bar{\alpha} > 1$, i.e.
the Landau splitting is  smaller than the splitting induced by the
SO interaction, one has $\alpha_{\kappa \tilde{\kappa}} \to
\tfrac{1}{2}$.  

Due to the SO interaction the Landau levels of different helicity
cross as seen in Fig.~\ref{fig_one}b and c. To demonstrate the 
influence of scattering around these crossing points we apply the
described extension of the SCBA to a two level system with different
spacings.  
The two states at the energies $E_{r}$, $E_{l}$ have different
helicity and evolve with increasing SO coupling from the spin 
{\it up}/{\it down} states with energy $E_{u}$ and $E_{d}$.

\begin{figure}[h!]
\includegraphics[width=0.8\columnwidth]{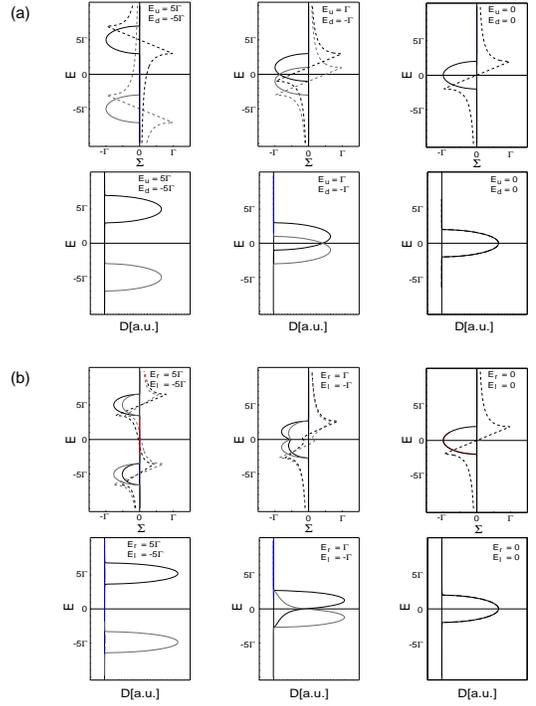}
\caption{\label{fig_two} Selfenergies $\Sigma(E)$ and density of
states for a two-level system without (a) and with (b) SO coupling for
different separation of the up-/down- or right-/left-eigenstates in
(a) and (b), respectively. In (b) a mixing coefficient $\alpha_{\kappa
\kappa} = 0.6$ at the energies $E_{r}$ and $E_{l}$ has been chosen.}
\end{figure}

In Fig.~\ref{fig_two}a the situation without 
SO interaction $\alpha_{\kappa \tilde{\kappa}} = \delta_{\kappa
\tilde{\kappa}}$ is shown. For decreasing energy difference $E_{u} -
E_{d}$ the form of  the selfenergy, calculated from
Eq.~(\ref{eqn_sigma_tautau}), and 
the density of states, from Eq.~(\ref{eqn_dos_tau}), keeps unchanged. 
In contrast, in the case of SO coupled states (Fig.~\ref{fig_two}b, with $\alpha_{\kappa \kappa} = 0.6$) the
impurity scattering adds to the selfenergy a contribution at the
neighbouring state with opposite  helicity. If the
states are not degenerate, both width and height of the imaginary part
of the selfenergy are reduced as compared  
with Fig.~\ref{fig_two}a, while for the
degenerate levels the selfenergy and the density of states are
identical with the situation of pure spin states.
For strongly SO coupled states $\alpha_{\kappa \kappa} \to \tfrac{1}{2}$
the height and width of the imaginary part of the selfenergy are
reduced by $1/\sqrt{2}$, as can be seen from
Eq.~(\ref{eqn_sigma_tautau}).  
As these parameters are inversely proportional to the relaxation time we
conclude, that the scattering time is increased by the SO coupling,
away from the crossing points. 

To summarize, the scattering strength depends on the level separation,
i.e. on  the
distance from crossing points in Fig.~\ref{fig_one}. As this distance 
varies with the magnetic field we expect a modulation of the
scattering strength with the magnetic field.


\section{Conductivity}
\label{sec_conductivity}

\begin{figure}[h!]
\includegraphics[width=0.8\columnwidth]{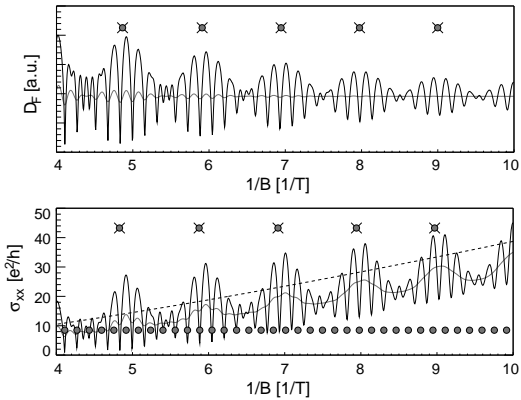}
\caption{\label{fig_three} Thermodynamic density of states  at the
Fermi energy  
$D_{\mathrm F}$ and longitudinal  
conductivity $\sigma_{xx}$ for the parameters of
Fig.~\ref{fig_one},  electron density $n_{s}= 3.0 \times 10^{15} \,$m$^{-2}$ and mobility
$\mu = 50\,$m$^{2}/$Vs at two
temperatures $T=1.0\,$K (black solid) and $3.0\,$K (grey solid). The
grey circless mark the expected position of  maxima of the SdH
oscillations without SO coupling; Landau level crossings due to the SO 
coupling are indicated by crosses. The dashed line in the plot of the
longitudinal conductivity is the classical limit 
of the SCBA when the Landau levels are decoupled.}
\end{figure}

\begin{figure}[h!]
\includegraphics[width=0.8\columnwidth]{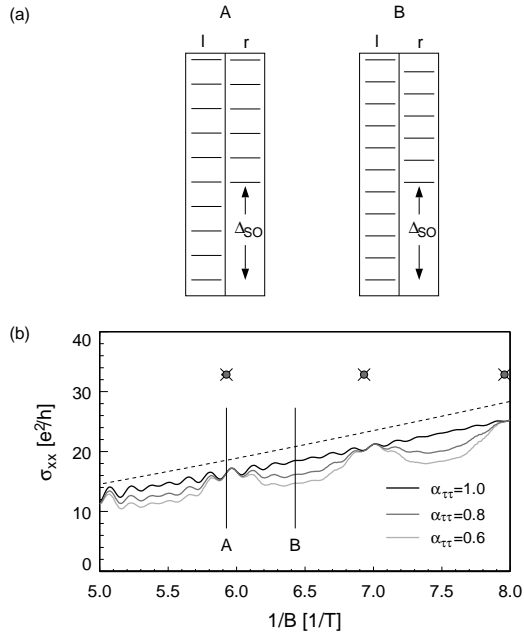}
\caption{\label{fig_four} (a) Sketch of Landau level spectrum with (A) 
and without (B) degeneracy of states with different helicity. (b)
Longitudinal magnetoconductivity in dependence of the strengh of the
SO coupling expressed by the parameter $\alpha_{\kappa
  \kappa}$ for high temperature $T=3.0\,$K.
The parameters are those of Fig.~\ref{fig_three} and  the dashed line
is again  the classical limit 
of the SCBA when the Landau levels are decoupled.}
\end{figure}

Based on the exact eigenstates of the constant SO coupling model
we evaluate the Kubo formula
\begin{equation}
\label{eqn_cond_mumu}
\sigma_{\mu \mu} =  \frac{e^{2} \pi \hbar}{L_{x} L_{y}} \int dE (-\frac{df_{0}}{dE}) \sum_{\alpha \alpha'}
\left| \left\langle \alpha | v_{\mu} | \alpha' \right\rangle \right|^{2}
A_{\alpha}\, A_{\alpha'} 
\end{equation} 
to calculate the conductivity. Here the spectral function $A_{\alpha}$
includes the impurity scattering in SCBA as described in
Sec.~\ref{third_section}  and the Fermi distribution function yields
the 
temperature average. 
The velocity is given by the equation $i \hbar v_{\mu} = [x_{\mu},H]$, 
which results in
\begin{align}
v_{x} = \tfrac{1}{m^{*}} \left( \pi_{x} + \tfrac{\alpha_{z}}{\hbar}
\sigma_{y} \right) \;\;\;\;\; \text{and} \;\;\;\;\;\; 
v_{y} = \tfrac{1}{m^{*}} \left( \pi_{y} -
\tfrac{\alpha_{z}}{\hbar} \sigma_{x} \right).
\end{align}
Besides $\sigma_{\mu \mu}, \mu = x, y$  we calculate also the thermodynamic 
density of states at the Fermi energy 
$D_{\mathrm{F}} = \int dE (-\frac{df_{0}}{dE}) D(E)$. In  
Fig.~\ref{fig_three} both quantities are shown for the two temperatures 
$T=1\,$K and $T=3\,$K with the Fermi energy determined from the
constant electron density $n_{s}$. For comparison the 
classical high temperature limit of the SCBA for decoupled Landau
levels is given by the dashed line.
The effect of spin-orbit coupling is seen in $D_{F}$
as the beating  pattern, well-known 
from measured SdH oscillations \cite{nitt97,heid98,schae98,hu99},
while in $\sigma_{xx}$ it causes an 
additional 
modulation, which survives even at a higher temperature when the
SdH oscillations are damped out. The period of this modulation is
determined by the crossing of Landau levels (marked in
Fig.~\ref{fig_three}) induced by the spin-orbit 
coupling. 
The self-energy enters differently into $D_{\mathrm{F}}$ and
$\sigma_{xx}$. In $D_{\mathrm{F}}$ it leads to a modulated broadening
of the Landau levels which is washed out in the high-temperature
limit, while in $\sigma_{xx}$ it acts in addition as a scattering time 
whose dependence on the level separation remains even at higher
temperatures. This is seen by comparing with the classical SCBA limit
for decoupled Landau levels (dashed lines): the Kubo formula
(\ref{eqn_cond_mumu}) yields a magnetoconductivity which is reduced
away from the crossing points due to suppression of impurity
scattering. (Note, that we are in the limit $\omega_{\mathrm{c}} \tau
\gg 1$, where the Drude conductivity is proportional to $1/\tau$.) 
To demonstrate the effect of scattering between states 
of different helicity in our extension of the SCBA, we show  
in Fig.~\ref{fig_four} the dependence of the conductivity at $T=3\,$K on
the mixing coefficient $\alpha_{\kappa \tilde{\kappa}}$ by keeping the
energy spectrum with spin-orbit splitting unchanged but varying
$\alpha_{\kappa \kappa}$, which otherwise is given by the coefficients
$c_{\kappa}^{\sigma}$.   
Neglecting the scattering between states of different helicity      
$\alpha_{\kappa  \kappa} = 1$, 
the crossing of Landau levels has no effect on the conductivity. For
dominant spin-orbit coupling,
 $\alpha_{\kappa \kappa} \to \tfrac{1}{2}$, we see that the
conductivity is reduced between crossing points of Landau levels.
We can distinguish two situations (top of Fig.~\ref{fig_four}): The
states with different helicity are degenerate  (A) and
the scattering efficiency keeps unchanged; When
{\it right}/{\it left}-states are not degenerate  (B) the conductivity
depends on $\alpha_{\kappa \tilde{\kappa}}$.
\begin{figure}[h!]
\includegraphics[width=0.7\columnwidth]{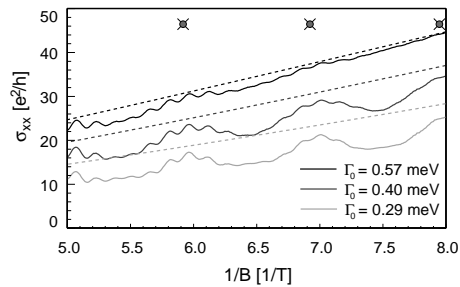}
\caption{\label{fig_five} Longitudinal magnetoconductivity at fixed
$\alpha_{rr}=0.6$ but with different $\Gamma_{0}$ at $T=3.0\,$K. The
other parameters are those of Fig.~\ref{fig_three}. The dashed lines mark the classical limit of the SCBA with decoupled Landau levels.}
\end{figure}
In Fig.~\ref{fig_five}, we have varied the strength of the 
impurity scattering by changing the parameter $\Gamma$. At the  classical limit of the SCBA
with decoupled Landau levels (dashed lines), we see that the 
conductivity rises with increasing $\Gamma$. The
modulation of the conductivity due to the crossing of Landau levels
decreases, because the limit of decoupled Landau levels cannot be
reached for large enough $\Gamma$.   

We have shown by a quantum mechanical calculation, that the SO
coupling  causes the expected beats of the SdH oscillations. In
addition our results exhibit a new modulation of the
magnetoconductivity which can be ascribed to a modification of
impurity scattering in the presence of SO coupling. This could help in
experiments to distinguish the 
effect of the SO coupling from that of inhomogeneous electron
densities which was invoked in Ref.~\cite{bros99} to explain beatings
in the SdH oscillations.

\section{Magnetotransport in  lateral superlattices with spin-orbit
coupling} 
\label{fifth_section}

\begin{figure}[h!]
\includegraphics[width=0.8\columnwidth]{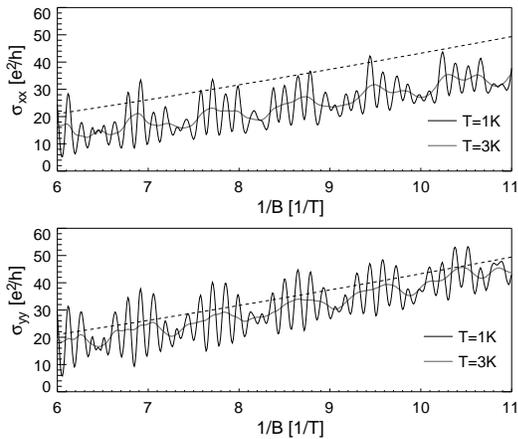}
\caption{\label{fig_six} Longitudinal magnetoconductivities
$\sigma_{xx}$ and $\sigma_{yy}$ for a lateral superlattice  with
uni-directional modulation in $x$-direction. (modulation amplitude
$V_{0}=3\,$meV, lattice constant $a=75\,$nm, Fermi energy
$E_{\mathrm{F}}=30\,$meV, mobility  $\mu=50\,$m$^{2}/$Vs and InAs
effective mass $m^{*}=0.0229\,m_{e}$. The
SO coupling is $\alpha_{z}=2.0\times10^{-11}\,$eVm and the mixing coefficient $\alpha_{\kappa \kappa}=0.5$. }
\end{figure}

Lateral semiconductor superlattices have proven to be well suited  for
studying the physics of Bloch electrons in artificial
periodic systems \cite{weis91}.  A  variety of oscillations have been
observed in 
magnetotransport experiments for systems, in which the lattice
constant $a$ is comparable with
achievable magnetic lengths $\lambda_{c}=\sqrt{\frac{\hbar}{e B}}$ and
both being much smaller (at low temperature) than the carrier
mean-free-path. After having studied the SdH oscillations in the
previous section we focus here on the influence of the Rashba SO
interaction on the various magnetoconductivity oscillations due to
the periodic modulation mentioned in the introduction.   
For the lateral superlattice with Rashba SO interaction the
Hamiltonian is given by 
\begin{equation}
\label{eqn_ham_so_vmod}
H =
\begin{pmatrix}
\frac{1}{2 m^{*}} (\pi_{x}^2 + \pi_{y}^{2}) &
\frac{\alpha_{z}}{\hbar} (\pi_{x} + i \pi_{y})\\
\frac{\alpha_{z}}{\hbar} (\pi_{x} - i \pi_{y}) &
\frac{1}{2 m^{*}} (\pi_{x}^2 + \pi_{y}^{2})
\end{pmatrix} + 
\begin{pmatrix}
V(x,y) & 0 \\
0      & V(x,y)
\end{pmatrix}.
\end{equation}
We consider here a periodic modulation in $x$ direction described by 
\begin{equation}
\label{eqn_cos_pot2}
V(x) = \frac{V_{0}}{2} \cos( \frac{2 \pi}{a} x).
\end{equation}
It removes the degeneracy of the Landau levels and is taken into
account in evaluating the Kubo formula together with the
spin-conserving impurity scattering in the extension of the SCBA as
for the homogeneous 2DES. In Fig.~\ref{fig_six} the longitudinal
magnetoconductivities $\sigma_{xx}$ and $\sigma_{yy}$, in the direction
of the modulation and perpendicular to it, respectively, are depicted
for potential parameters $V_{0}=3\,$meV, $a=75\,$nm, and two
temperatures. The Fermi energy was fixed to $E_{F}=30\,$meV, much
larger  
than the amplitude of the periodic potential, which defines the weak
modulation case. The mixing coefficient was fixed to the maximum value
$\alpha_{\kappa \kappa} = \tfrac{1}{2}$. Different types of
oscillations can be identified of which at the higher temperature
($3\,$K) only the commensurability oscillations and the modulation due 
to SO coupling survive. The
periods  (seen in the $1\,$K traces) can be quantified in
simplified models. Following Onsager \cite{onsa31} the SdH periods are 
\begin{equation}
\label{eqn_d1db_sdhpm}
\Delta_{1/B}^{\mathrm{SdH:+/-}} = \frac{e \hbar}{m^{*} (E_{\mathrm{F}}
\pm 
\tfrac{1}{2} \Delta_{\mathrm{SO}})},
\end{equation}
where the appearance of two Fermi contours due to the spin-splitting
is accounted for \cite{kepp02}. The period of the commensurability
oscillations in given by \cite{maga94}  
\begin{equation}
\label{eqn_d1db_copm}
\Delta_{1/B}^{\mathrm{CO:+/-}} = \frac{e a}{2 \sqrt{2 m^{*}
(E_{\mathrm{F}}\pm \tfrac{1}{2}\Delta_{\mathrm{SO}})}}.
\end{equation}
where again the spin-split Fermi contours are considered. The periods
reflecting the formation of the miniband structure due to the periodic 
potential \cite{albr98,lang00,deut01} are quantified by
\begin{equation}
\label{eqn_onsager}
\Delta_{1/B} = \frac{2 \pi e}{\hbar} \frac{1}{A},
\end{equation}
where $A$ is the cross section of the modified Fermi contour given in
Ref. \cite{lang00}. Again the spin-splitting due to SO coupling is to
be considered and Eq. (\ref{eqn_onsager}) yield two periods for the
Fermi cross sections
\begin{align}
\label{eqn_a_1dpm}
& A^{\mathrm{1D:+/-}} = 
\frac{2 m^{*}(E_{\mathrm{F}}\pm \tfrac{1}{2} \Delta_{\mathrm{SO}}) \pi}
{\hbar^{2}} 
 -\\
&- 2  \left(  
\frac{\pi}{a} 
\right)
\sqrt{
\frac{2 m^{*} (E_{\mathrm{F}}\pm  \tfrac{1}{2}
\Delta_{\mathrm{SO}})}{\hbar^{2}} 
 - \left(\frac{\pi}{a}\right)^{2}
} 
- \nonumber \\
& - 
\frac{4 m^{*} (E_{\mathrm{F}}\pm\tfrac{1}{2} \Delta_{\mathrm{SO}} )
}{\hbar^{2}}  
\arcsin{
\frac{\pi \hbar}{a 
\sqrt{2m^{*} (E_{\mathrm{F}}\pm  \tfrac{1}{2} \Delta_{\mathrm{SO}})
}}}.
\end{align}
Finally we may conclude from the eigenvalues of
Eq.~(\ref{eqn_eigenvaluesrl}) for $n \gg 1$ a period of 
\begin{equation}
\label{eqn_sod1db}
\Delta_{1/B}^{\mathrm{SO}} = \frac{e \hbar}{m^{*} \Delta_{\mathrm{SO}}} 
\end{equation}
for the modulation connected  with the crossing points of the Landau
levels at the Fermi energy.  

\begin{figure}[h!]
\includegraphics[width=0.7\columnwidth]{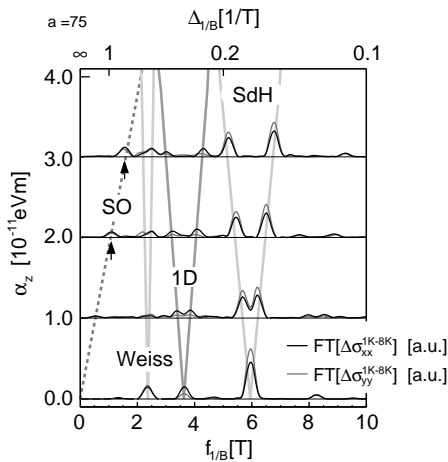}
\caption{\label{fig_seven} The absolute value of the Fourier spectrum
of the difference of the conductivities at $1\,$K and $8\,$K analogous
to Fig.~\ref{fig_six} is shown in 
dependence on $\alpha_{z}$. The arrows mark the contribution due to the
crossing of Landau levels. The calculated  Fourier spectrum is
compared with the models of
Eqs.~(\ref{eqn_d1db_sdhpm}),(\ref{eqn_d1db_copm}), (\ref{eqn_onsager}) and (\ref{eqn_sod1db}).}

\end{figure}

\begin{figure}[h!]
\includegraphics[width=0.7\columnwidth]{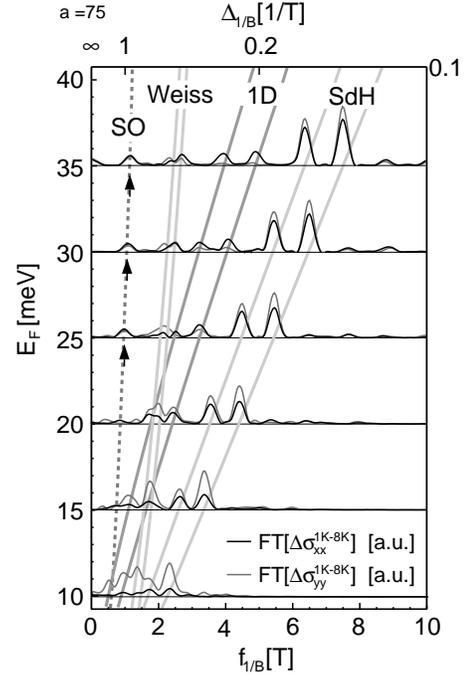}
\caption{\label{fig_eight}  The absolute value of the Fourier spectrum
of the difference of the conductivities at $1\,$K and $8\,$K analogous
to Fig.~\ref{fig_six} is shown in
dependence on the Fermi energy $E_{\mathrm{F}}$. The arrows mark the
contribution due to the  crossing of Landau levels. Comparison with
Eqs.~(\ref{eqn_d1db_sdhpm}),(\ref{eqn_d1db_copm}), (\ref{eqn_onsager})
and (\ref{eqn_sod1db}) is shown by straight lines.}
\end{figure}

In order to analyze all these oscillations we take the power spectrum
of the differences $\Delta \sigma_{\mu \mu}$, $\mu=x,y$, of the
calculated longitudinal conductivities at $1$ and $8\,$K. It is shown
in dependence on the strength of the SO coupling (Fig.~\ref{fig_seven})
and on the Fermi energy $E_{\mathrm{F}}$ (Fig.~\ref{fig_eight})
together with the periods predicted from
Eqs. (\ref{eqn_d1db_sdhpm})-(\ref{eqn_sod1db}). In
Fig.~\ref{fig_seven} we see that
with increasing SO coupling the frequencies of all (but one)
resolved 1/B periodic oscillations split and the maxima of the
Fourier transform follow the predictions of the simplified models.
This is not the case for the peaks showing up at the lowest $f_{1/B}$
values. These peaks, being due to the modulation resulting from our
extension of the SCBA (Sec. \ref{third_section}), do not split and
follow the analytic expressions of Eq.~(\ref{eqn_sod1db}).
In Fig.~\ref{fig_eight} the oscillation period is shown in dependence
on the Fermi energy. 
Again the
maxima of our full quantum-mechanical calculation follow the
predictions of the simplified models of
Eqs. (\ref{eqn_d1db_sdhpm})-(\ref{eqn_sod1db}).

\section{Summary}
We have calculated the magnetoconductivity for a homogenous 2DES with
spin-orbit interaction including a non-trivial extension of the SCBA
which takes into account the spin-degree of freedom. The crossing of
Landau levels with different helicity, which is induced
by the spin-orbit interaction, manifests itself in
an additional modulation of the conductivity. It is due to a
modification of impurity scattering in the presence of SO
coupling. This signature could be used - besides the beating of SdH
oscillations - for the experimental verification of SO coupling.

Further we have investigated the influence of SO interaction on the
magnetoconductivity oscillations in 1D lateral superlattices. The
prediction of the frequencies by simple models and the numerical
calculations are in agreement and should motivate an experimental 
verification.

\begin{acknowledgments}
This work has been supported by the DFG via Forschergruppe~370 {\em
Ferromagnet-Halbleiter-Nanostrukturen}.
\end{acknowledgments}


\vspace{2ex}





\begin{thebibliography}{9}


\bibitem{rash60} \'{E}.I~Rashba, Sov.~Phys.~Sol.~Stat. {\bf 2}, 1109 (1960).

\bibitem{bych84}  Yu.A.~Bychkov and \'{E}.I.~Rashba, JETP Lett. {\bf 39}, 78 (1984).

\bibitem{wolf01} S.A.~Wolf, D.D.~Awschalom, R.A.~Buhrman,
  J.M.~Daughton, S.~von~Moln{\'a}r, M.L.~Roukes, A.Y.~Chtchelkanova,
  D.M.~Treger, Science {\bf 294}, 1488 (2001). 

\bibitem{dres55}  G.~Dresselhaus, Phys.~Rev. {\bf 100}, 580
  (1955).

\bibitem{juss} L.~Wissinger, U.~R\"{o}ssler, R.~Winkler, B.~Jusserand, 
and D.~Richards, Phys. Rev. B {\bf 58}, 15375 (1998).


\bibitem{pikus_pikus} F.G.~Pikus and G.E.~Pikus, Phys Rev. B {\bf 51}, 
16928 (1995).

\bibitem{schie02} Ch.~Schierholz, R.~K\"{u}rsten, G.~Meier,
T.~Matsuyama, and U.~Merkt, phys.~stat.~sol.~(b) {\bf 233}, 436 (2002).

\bibitem{titkov} G.E.~Pikus, A.N.~Titkov in: Optical Orientation,
edited by F.~Meier and B.P.~Zakharchenya, North-Holland, Amsterdam
(1984). 

\bibitem{dp} M.I.~Dyakonov and V.I.~Perel, Sov. Phys. Solid State
{\bf 13}, 3023 (1972).

\bibitem{golub} N.S.~Averkiev and L.E.~Golub, Phys. Rev. B {\bf 60},
15582 (1999).

\bibitem{jk} J.~Kainz, U.~R\"{o}ssler, and R.~Winkler, cond-mat/0304017
(2003). 

\bibitem{ganichev} S.D.~Ganichev, S.N.~Danilov, V.V.~Bel'kov,
E.L.~Ivchenko, M.~Bichler, W.~Wegscheider, D.~Weiss, and W.~Prettl,
Phys. Rev. Lett. {\bf 88}, 057401  (2002). 

\bibitem{lomm88} G.~Lommer, F.~Malcher and U.~R\"{o}ssler,
Phys.~Rev.~Lett. {\bf 60}, 728 (1988). 


\bibitem{nitt97} J.~Nitta,T.~Akazaki, H.~Takayanagi, T.~Enoki,
Phys. Rev. Lett. {\bf 78}, 1335 (1997). 

\bibitem{heid98} J.P.~Heida, B.J.~van~Wees, J.J.~Kuipers, and T.M.~Klapwijk, G.~Borghs, Phys. Rev. B {\bf 57}, 11911 (1998).

\bibitem{schae98} Th.~Sch\"{a}pers, G.~Engels, J.~Lange,
  Th.~Klocke, M.~Hollfelder, and H.~L\"{u}th, J.~Appl.~Phys. {\bf 98}, 4324 (1998).

\bibitem{hu99} C.-M.~Hu,J.~Nitta, T.~Akazaki, H.~Takayanagai, J.~Osaka, P.~Pfeffer, W.~Zawadzki, Phys.~Rev.~B {\bf 60}, 7736 (1999).

\bibitem{wink00} R.~Winkler, S.J.~Papadakis, E.P.~De~Poortere, and
M.~Shayegan, Phys.~Rev.~Lett.~{\bf 84}, 713 (2000).

\bibitem{kepp02} S.~Keppeler, R.~Winkler, Phys.~Rev.~Lett.~{\bf 88},
046401 (2002).

\bibitem{bros99} S. Brosig, K. Ensslin, R. J. Warburton, C. Nguyen,
B. Brar, M. Thomas, H. Kroemer, Phys.~Rev.~B {\bf 60}, R13989
(1999). 

\bibitem{weis91} D.~Weiss, Adv.~Sol.~Stat.~Phys. {\bf 31},341 (1991).

\bibitem{weis89_wink89} D.~Weiss, K.~von~Klitzing, K.~Ploog, and G.~Weimann, Europhys.~Lett. {\bf 8}, 179 (1989); 
R.W.~Winkler, J.P.~Kotthaus, and K.~Ploog, Phys.~Rev.~Lett. {\bf 62}, 1177 (1989).


\bibitem{albr99} C.~Albrecht, et~al., Phys.~Rev.~Lett. {\bf 83}, 2234
(1999). 


\bibitem{deut01}  R.A.~Deutschmann,  W.~Wegscheider,
M.~Rother, M.~Bichler, G.~Abstreiter, C.~Albrecht, J.H.~Smet, 
Phys.~Rev.~Lett. {\bf 86}, 1857 (2001).


\bibitem{lang02} M.~Langenbuch, M.~Suhrke, U.~R\"{o}ssler, Proceedings
2nd PASPS W\"{u}rzburg, accepted for publication in Journal of
Superconductivity (2002). 

\bibitem{maga94} L.L.~Magarill, Superlattices and
Microstructures {\bf 16}, 257 (1994). 

\bibitem{weis89} D.~Weiss, K.~von Klitzing, K.~Ploog, and G.~Weimann, Europhys.~Lett. {\bf 8}, 179 (1989); \\
R.W.~Winkler, J.P.~Kotthaus and K.~Ploog,
Phys.~Rev.~Lett. \textbf{62}, 1177 (1989).

\bibitem{ando82} T.~Ando, A.B.~Fowler and F.~Stern, Reviews of Modern Physics {\bf 54}, 437 (1982).


\bibitem{gerh75} R.R.~Gerhardts, Zeitschrift f\"{u}r Physik B {\bf 22}, 327 (1975).




\bibitem{albr98} C.~Albrecht, J.H.~Smet, D.~Weiss,
K.~von~Klitzing, V.~Umansky, H.~Schweizer, Physica B {\bf 249-251},914
(1998).  


\bibitem{lang00} M.~Langenbuch, R.~Hennig, M.~Suhrke,
  U.~R\"{o}ssler, C.~Albrecht, J.H.~Smet, D.~Weiss, Physica E {\bf 6}, 565 (2000).

\bibitem{onsa31} L.~Onsager, Phys. Rev. {\bf 37}, 405 (1931).

 

\end{thebibliography}
\end{document}